\documentclass[a4paper,twocolumn,superscriptaddress,11pt,accepted=2019-08-19]{quantumarticle}
\pdfoutput=1
\usepackage[utf8]{inputenc}
\usepackage[english]{babel}
\usepackage[T1]{fontenc}
\usepackage{amsmath}
\usepackage{dsfont}
\usepackage{hyperref}
\usepackage[numbers,sort&compress]{natbib}

\usepackage{tikz}
\usepackage{lipsum}

\begin{document}

\title{Quantum correlations for anonymous metrology}
\date{\today}

\author{A.~J.~Paige}
\email{a.paige16@imperial.ac.uk}
\affiliation{QOLS, Blackett Laboratory, Imperial College London, London SW7 2AZ, UK.}
\author{Benjamin Yadin}
\email{benjamin.yadin@gmail.com}
\affiliation{Department of Atomic and Laser Physics, Clarendon Laboratory, University of Oxford}
\author{M.~S.~Kim}
\affiliation{QOLS, Blackett Laboratory, Imperial College London, London SW7 2AZ, UK.}

\maketitle

\begin{abstract}
	We introduce the task of anonymous metrology, in which a physical parameter of an object may be determined without revealing the object's location. Alice and Bob share a correlated quantum state, with which one of them probes the object. Upon receipt of the quantum state, Charlie is then able to estimate the parameter without knowing who possesses the object. We show that quantum correlations are resources for this task when Alice and Bob do not trust the devices in their labs. The anonymous metrology protocol moreover distinguishes different kinds of quantum correlations according to the level of desired security: discord is needed when the source of states is trustworthy, otherwise entanglement is necessary.
\end{abstract}

\section{Introduction}
Quantum correlations can provide significant advantages over classical resources in performing non-local tasks. Teleportation of an unknown quantum state~\cite{Bennett1993}, super dense coding~\cite{Bennett1992}, and quantum key distribution~\cite{Ekert1991}, are a few examples amongst other contexts~\cite{Horodecki2009,Modi2012,Adesso2016,Bera2018}. This has notably included their relevance to the power of quantum computation~\cite{Jozsa2002,Vidal2003,Datta2008,Dakic2010,DATTA2011}, their uses in metrology~\cite{Giovannetti2006,Huang2016,Modi2011,Girolami2013,Girolami2014}, and their roles in non-local information tasks like quantum state merging~\cite{Horodecki2005,Cavalcanti2011,Madhok2011}. Quantum correlations can also hide information non-locally~\cite{DiVincenzo2002,DiVincenzo2003,Christandl2005,Bouda2007,Brassard2007}, with recent studies investigating non-locally hiding computations~\cite{Shahandeh2017}, and when it is possible to mask quantum information~\cite{Modi2018}.

In this work, we introduce the task of anonymous metrology, which involves encoding an initially unknown continuous parameter in a state whilst hiding where the encoding happened. We identify the quantum states that enable the task and separately treat the two cases of having a trustworthy or untrustworthy source of states. We term the resourceful states as weakly anonymous (WA) and strongly anonymous (SA) respectively, and give physical intuition for the distinction by demonstrating how SA states allow the encoding's location to be not just hidden but quantum mechanically delocalised.

We derive general forms for the WA and SA states, using modes of translational asymmetry~\cite{Marvian2016,Marvian2013} for the former, and for the latter showing equivalence to the entangled ``maximally correlated states''~\cite{Rains1999} extended by degeneracy. We then determine the nature of their quantum correlations. In general, quantum correlations exist in different ``strengths'', from discord~\cite{Modi2012} to full Bell nonlocality~\cite{Brunner2014}, and understanding their respective utilities remains to be fully explored. Whilst there are several works that demonstrate an operational use for discord~\cite{Weedbrook2016,Cavalcanti2011,Girolami2014,Modi2011}, our results additionally reveal an operational distinction between different types of quantum correlations. We find that WA states require a form of discord that we term \emph{aligned discord}, while SA states require a stronger type of correlation, correspondingly termed \emph{aligned entanglement}.

\begin{figure}[t]
  \centering
  \includegraphics[scale=0.36]{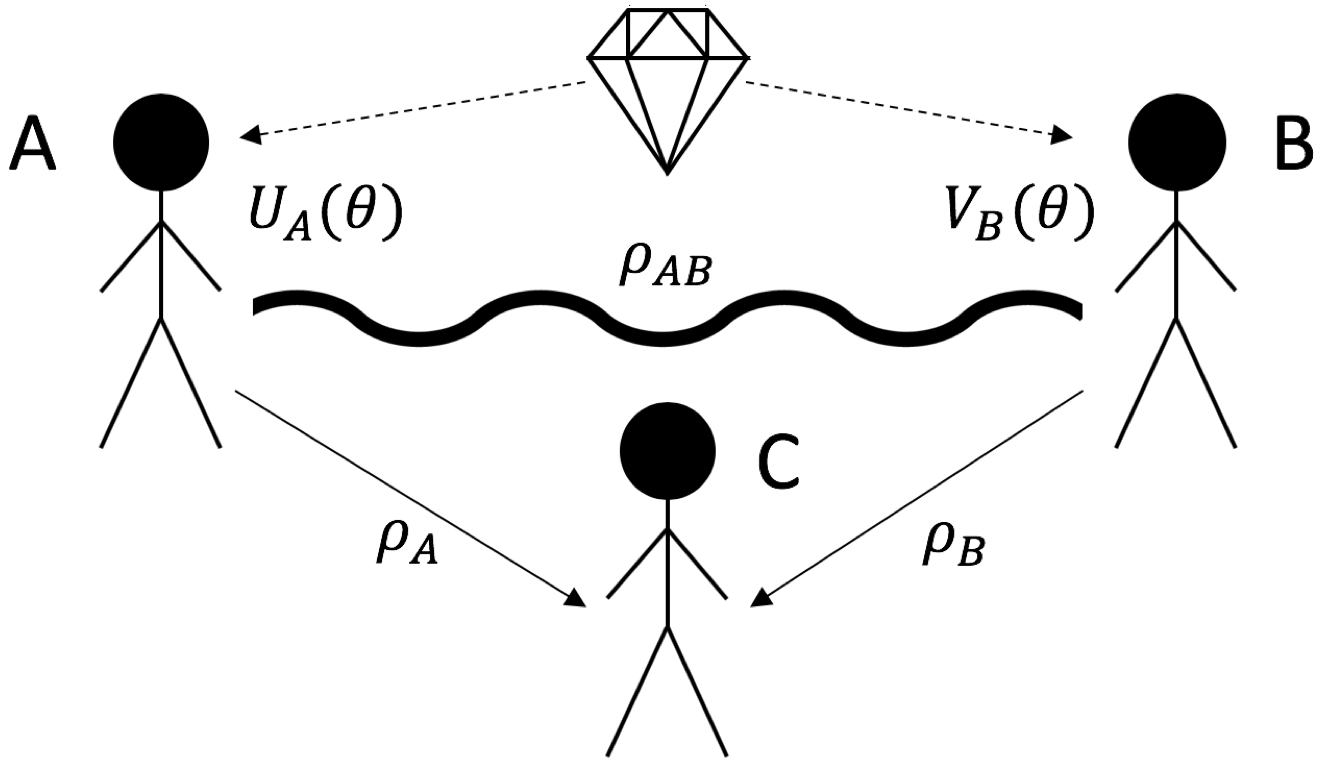}
  \caption{Illustration of the anonymous metrology task. Initially A and B share the state $\rho_{AB},$ and one of them is given the system to hide. They engineer $U_{A},$ or $V_{B},$ to encode $\theta$ into the shared state, then both halves are sent to C.}
  \label{AnonymousMetFig}
\end{figure}

\section{Defining anonymous metrology}
We introduce the task of anonymous metrology with an example, illustrated in Fig.~\ref{AnonymousMetFig}. Alice and Bob are in spatially separated laboratories, and one of them receives a system, the location of which they must keep hidden (e.g. some valuable diamond). Charlie wants them to probe it with a (finite-dimensional) quantum system to give him information about some initially unknown continuous parameter $\theta$ (such as a refractive index).

We make the following assumptions:
\begin{itemize}
	\item If the object is in Alice's lab, it interacts with her system such that the latter undergoes a unitary transformation $U_A(\theta) = e^{-i\theta H_A}$ with some parameter-independent Hamiltonian $H_A$. (Otherwise, Alice's system is unchanged.) Similarly, Bob's system undergoes $V_B(\theta) = e^{-i\theta G_B}$.
	\item Alice and Bob may use classical communication freely, but this is not secure. Charlie is also allowed to know the set-up of their labs.
	\item Their devices are unsecure, in that any measurement outcomes obtained are assumed to be available to Charlie. Note that they may implement any parameter-independent local unitaries without loss of security. Hence it is possible to effectively change the eigenbasis of $H_A,G_B$ arbitrarily.
	
\end{itemize}

The task is to enable Charlie to estimate $\theta$, but prevent him from learning where the hidden system is, i.e. the one who actually encoded $\theta$ into the quantum system must remain anonymous.

At first glance this task may seem impossible, since $\theta$ is initially unknown and Alice and Bob can only learn its value via untrusted local measurements. With Charlie accessing their measurement results we might fear that he will always be able to determine where the encoding is happening and thus the system's location. However it turns out that quantum resources allow them to succeed in this task.

We formalize this statement later, but we shall start by illustrating with a simple example using an entangled state. We allow Alice and Bob to request copies of a bipartite quantum state $\rho_{AB}.$ Consider the situation when they choose the Bell state $|\psi^{+}\rangle_{AB}=\frac{1}{\sqrt 2}(|00\rangle_{AB}+|11\rangle_{AB}).$ Alice can apply $U_A(\theta)=e^{-i\theta |1\rangle_A\langle 1|},$ to produce $|\psi(\theta)\rangle_{AB}=\frac{1}{\sqrt 2}(|00\rangle_{AB}+e^{-i\theta}|11\rangle_{AB}),$ but similarly Bob can apply $V_B(\theta)=e^{-i\theta |1\rangle_B\langle 1|},$ and for all $\theta$ he produces the same state. This indicates the solution to their problem. The one who has the hidden system interacts their half of $\rho_{AB}$ with it, to realize the relevant encoding unitary (they may need a rescaling such that $\theta\in [0,2\pi)$), and then they both send their halves of $\rho_{AB}$ to Charlie. Given multiple copies Charlie can determine $\theta$ to arbitrary precision but cannot tell if it was $U_A$ or $V_B$ that changed the state, so cannot learn the system's location\footnote{Note instead of sending states to Charlie they could perform a set of informationally complete local measurements, tomographically reconstruct the state and extract $\theta$ from it, but this is unnecessarily complicated and would generally be less efficient for learning $\theta.$}. Clearly the Bell state is a resource for this task, but we shall show that a number of quantum states are, and in some cases they are not entangled.

\subsection{Anonymity and encoding conditions}
For anonymous metrology there are two relevant sets of useful states, and the choice between them depends on who provides the states for Alice and Bob.

First consider the situation where a trustworthy fourth party, who will never share information with Charlie, is sending quantum states to them. Alice and Bob should request copies of a state $\rho_{AB},$ that satisfies two conditions. First that they can find continuously parametrised unitaries $U_{A}(\theta),V_{B}(\theta)$  such that, for $\theta$ in some interval,
\begin{equation} \label{WACondition}
U_A(\theta)\rho_{AB}U^\dagger_A(\theta)=V_B(\theta)\rho_{AB}V^\dagger_B(\theta).
\end{equation}
This is termed the \emph{weak anonymity condition}, since it ensures that for a given parameter the same state is produced no matter who encoded it. The second condition is that from the same interval, different phases produce different states, i.e. given $\theta\neq\phi$ we have
\begin{equation} \label{EncodingCondition}
 U_A(\theta)\rho_{AB}U^\dagger_A(\theta)\neq U_A(\phi)\rho_{AB}U^\dagger_A(\phi).
\end{equation}
We term this the \emph{encoding condition} since it ensures that different parameters are mapped to different states, so in principle Charlie learns at least some information about the parameter.

If however it is Charlie who is sending the states (or the fourth party is untrustworthy) then anonymity with these states is not assured. The most dangerous situation is when Charlie holds a third system and knows the pure state $|\psi\rangle_{ABC},$ but Alice and Bob only know the state $\rho_{AB}=\text{Tr}_{C}(|\psi\rangle_{ABC}\langle\psi |).$ They can verify that they have been sent copies of $\rho_{AB}$ by using a subset of the states they receive to perform metrology, but by doing this they cannot learn what Charlie holds. We also note that using a finite number of copies for metrology will inevitably lead to some finite error, and we address this later in section \ref{sec:NonIdealStates} where we show robustness for the protocol.

To maintain anonymity when Charlie could be holding a purification, they must be able to find unitaries such that $U_{A}(\theta)|\psi\rangle_{ABC}=V_{B}(\theta)|\psi\rangle_{ABC},$ up to an irrelevant global phase. We use this to derive a condition on the states $\rho_{AB}$ that they can choose. We expand using the Schmidt decomposition $|\psi\rangle_{ABC}=\sum_j \lambda_j |\phi_j\rangle_{AB}\otimes|\chi_j\rangle_C$ (in terms of an orthogonal product basis), and projecting onto $|\chi_j\rangle_C,$ we have $U_{A}(\theta)|\phi_j\rangle_{AB} = V_{B}(\theta)|\phi_j\rangle_{AB}.$ Writing $\rho_{AB}=\sum_{j}|\lambda_{j}|^2|\phi_j\rangle_{AB}\langle \phi_j|,$ and acting from the left with $U_A(\theta),$ we arrive at
\begin{equation}\label{SACondition}
U_A(\theta)\rho_{AB}=V_B(\theta)\rho_{AB}.
\end{equation}
This is termed the \emph{strong anonymity condition}. Reversing the argument is straightforward. Hence the state $|\psi\rangle_{ABC}$ has unitaries that satisfy $U_{A}(\theta)|\psi\rangle_{ABC}=V_{B}(\theta)|\psi\rangle_{ABC},$ if and only if $\rho_{AB}=\text{Tr}_{C}(|\psi\rangle_{ABC}\langle\psi |)$ has unitaries that satisfy Eq.~(\ref{SACondition}). It is clear that Eq.~(\ref{SACondition}) implies Eq.~(\ref{WACondition}), but not vice versa, so the condition (\ref{SACondition}) is stronger. With this we now formally establish appropriate terminology.

\vspace{1mm}
\noindent
\textbf{Definitions:} A state $\rho_{AB}$ is a \emph{weakly anonymous} (WA) state if there exist unitaries $U_{A}(\theta)=e^{-i\theta H_{A}},$ and $V_{B}(\theta)=e^{-i\theta G_{B}}$ that satisfy the conditions given by Eq.~(\ref{WACondition}) and (\ref{EncodingCondition}). The subset of these that also satisfy the condition of Eq.~(\ref{SACondition}) are \emph{strongly anonymous} (SA) states.
\vspace{1mm}

For pure states the WA and SA conditions coincide, and furthermore a pure state is WA/SA if and only if it is entangled. To prove sufficiency we use the Schmidt decomposition to write $|\psi\rangle_{AB}=\sum_j\lambda_j|\phi_j\rangle_A \otimes |\chi_j\rangle_B.$ Entangled states have a Schmidt number of at least two, so without loss of generality we take $\lambda_0\neq0$ and $\lambda_1\neq0.$ Now the unitaries $U_A(\theta)=e^{-i\theta |\phi_1\rangle\langle\phi_1|_A },$ and $V_B(\theta)=e^{-i\theta |\chi_1\rangle\langle\chi_1|_B },$ satisfy the conditions. Hence all pure entangled states are WA/SA. To prove entanglement is necessary consider a separable state $|\psi\rangle_{AB}=|\phi\rangle_A \otimes |\chi\rangle_B.$ The anonymity condition becomes $U_A(\theta)|\phi\rangle_A \otimes |\chi\rangle_B = |\phi\rangle_A \otimes V_B(\theta)|\chi\rangle_B.$ Applying the ket $\langle\phi|_A$ we get $\langle \phi |U_A(\theta)|\phi\rangle |\chi\rangle_B=V_B(\theta)|\chi\rangle_B,$ and projecting this equation onto itself we arrive at $|\langle \phi |U_A(\theta)|\phi\rangle|=1.$ Hence $U_A(\theta)$ only imparts an unobservable global phase, so violates the encoding condition. Therefore entanglement is necessary for pure states.

For mixed states things are more complicated and we address this shortly, after presenting a different non-local task that illustrates the true distinction between the WA and SA cases.

Hiding the location of a system from Charlie somewhat resembles hiding which-path information. Now consider Alice and Bob are tasked with measuring a system that is put in a superposition of going to Alice and Bob. Can they perform measurements without acquiring which-path information and thus without decohering the spatial superposition?

Formalizing this, we consider quantizing the path degree of freedom $P$ of the system to be measured; it is put into some superposition $a|L\rangle_{P} + b|R\rangle_{P}$ of going left to Alice and right to Bob. They probe the system with a shared state $\rho_{AB}$, described by the controlled unitary $W(\theta) = |L\rangle_P \langle L| \otimes U_A(\theta) + |R\rangle_P \langle R| \otimes V_B(\theta)$. Requiring the final state of $P$ to factor out unchanged, recovers the SA condition Eq.~(\ref{SACondition}) -- see Appendix \ref{appenNoDec} for details.

This emphasises the fact that the SA condition ensures no information exists on where the interaction took place. The measurement was delocalised by the correlations. We now move from operational considerations to investigate the form of the resourceful states.

\section{Form of useful states}
\subsection{Form of WA states}
In order to arrive at a form for the WA states, it is useful to employ modes of translational asymmetry~\cite{Marvian2016,Marvian2013}. Given our unitary action $\mathcal{U}_{A,\theta}(.)=U_A(\theta)(.)U_A^\dagger(\theta),$ a given state can be decomposed into modes as $\rho_{A}= \sum_{\omega}\rho_{A}^{(\omega)},$ where $\mathcal{U}_{A,\theta}(\rho_{A}^{(\omega)})=e^{i\omega\theta}\rho_{A}^{(\omega)}.$ This is akin to Fourier decomposition of a function. We can select out modes with the twirling superoperator
\begin{equation}
\mathcal{P}_{A}^{\omega}=\lim_{\theta_0 \rightarrow \infty}\frac{1}{2\theta_0}\int_{-\theta_0}^{\theta_0}d\theta e^{-i\omega \theta}\mathcal{U}_{A,\theta}.
\end{equation}
One can verify that $\mathcal{P}_{A}^{\omega}(\rho_{A})=\rho_{A}^{(\omega)}.$ We similarly define $\mathcal{P}_{B}^{\omega},$ using $\mathcal{V}_{B,\theta}(.)=V_B(\theta)(.)V_B^\dagger(\theta).$ The twirling operators satisfy the relation $\mathcal{U}_{A,\theta}\mathcal{P}_{A}^{\omega}=e^{i\omega\theta}\mathcal{P}_{A}^{\omega},$ and a completeness relation $\sum_{\omega}\mathcal{P}_{A}^{\omega} = \mathds{1}.$

We can rewrite the WA condition of Eq.~(\ref{WACondition}) in terms of superoperators by simply multiplying by $\frac{e^{-i\omega\theta}}{2\theta_0},$ integrating $\int_{-\theta_0}^{\theta_0} d\theta,$ and taking the limit $\theta_0 \rightarrow \infty,$ to get
\begin{equation}\label{WATwirling}
\mathcal{P}_{A}^{\omega}\rho_{AB}=\mathcal{P}_{B}^{\omega}\rho_{AB}.
\end{equation}
We prove the converse by acting on this equation with $\mathcal{U}_{A,\theta}\mathcal{V}_{B,\theta},$ to get $e^{i\omega\theta}\mathcal{V}_{B,\theta}\mathcal{P}_{A}^{\omega}\rho_{AB}=e^{i\omega\theta}\mathcal{U}_{A,\theta}\mathcal{P}_{B}^{\omega}\rho_{AB}.$ The $e^{i\omega\theta}$ terms cancel, and then summing over $\omega$ using the completeness relations we return to $\mathcal{V}_{B,\theta}\rho_{AB}=\mathcal{U}_{A,\theta}\rho_{AB}.$ Hence Eq (\ref{WATwirling}) captures the weak anonymity condition. Note that the encoding condition Eq.~(\ref{EncodingCondition}) becomes that there has to be some $\omega\neq 0$ for which $\mathcal{P}_{A}^{\omega}\rho_{AB}\neq0.$

From Eq.~(\ref{WATwirling}) we can explicitly write the form of the WA states. First we define
\begin{equation}
\rho^{(\omega_{1},\omega_{2})}_{AB}=\sum_{\substack{i,i',j,j',\\ E_{i'}=E_{i}+ \omega_{1},\\ E_{j'}=E_{j}+ \omega_{2}}}c_{ii'jj'} |i\rangle_A\langle i'|\otimes |j\rangle_B\langle j'|,
\end{equation}
where $H_{A}|i\rangle=E_{i}|i\rangle,$ with $H_{A}$ the Hamiltonian generator of $\mathcal{U}_{A,\theta},$ and similarly for $B.$ Then the WA states are of the form
\begin{equation}\label{WAStateForm}
\rho_{AB}=\sum_\omega \rho^{(\omega,\omega)}_{AB},
\end{equation}
where we require non-zero terms for $\omega\neq 0$ so that encoding is possible.

This shows that WA states are those with correlated modes of asymmetry, which indicates a connection with the resource of quantum coherence~\cite{Streltsov2017}. We can view WA states as having correlated coherence in the eigenbasis of the unitaries. There is a formal similarity with the correlated coherence defined in~\cite{Kraft2018,Tan2016,Ma2018,Sun2017}.

\subsection{Form of SA states}
We now derive the form of SA states. Here, working with modes of asymmetry is not as straightforward (see Appendix \ref{appenSATwirl} for more discussion), so we use a different approach.

Rearranging the anonymity condition of Eq.~(\ref{SACondition}) to $(U_A(\theta)-V_B(\theta))\rho_{AB}=0,$ and taking matrix elements in the eigenbasis of the local unitaries, we get $(u_{i}(\theta)-v_{i'}(\theta))\langle ii'|\rho_{AB}|jj'\rangle=0.$ The non-zero matrix elements are those for which $u_{i}(\theta)=v_{i'}(\theta).$ Initially it is simplest to consider the non-degenerate case so that $u_i(\theta)\neq u_j(\theta),\forall i\neq j$ and similarly for the $v_i$. With this the largest set of non-zero matrix elements is achieved by pairing every $u_{i}(\theta)$ with a $v_{i'}(\theta)$ such that $u_{i}(\theta)=v_{i'}(\theta).$ Since relabeling is physically irrelevant we can write the non-zero matrix elements as $\langle ii|\rho_{AB}|jj\rangle,$ and so we write the state as $\rho_{AB}=\sum_{i,j}\rho_{ij}|ii\rangle\langle jj|.$ This is the form of so-called ``maximally correlated" states~\cite{Rains1999}. Note that we need at least one non-zero off-diagonal $\rho_{ij}=\rho_{ji}^{*},$ to ensure that the encoding condition of Eq.~(\ref{EncodingCondition}) is satisfied.

We can lift the non-degeneracy restriction by introducing a new label, such that we write states that are degenerate under $U_A$ as $|i\lambda\rangle.$ We then write the form of SA states as
\begin{equation}\label{SAForm}
\rho_{AB}=\sum_{i,j,\lambda,\lambda',\mu,\mu'}\rho_{ij\lambda\lambda'\mu\mu'}|i\lambda , i\lambda'\rangle_{AB}\langle j\mu , j\mu'|.
\end{equation}
Hence the SA states are a generalisation of the ``maximally correlated" states, where we only include the entangled ones. Having established the forms of useful states (see Appendix \ref{appenMultiPart} for the generalisation to multipartite cases), we now discuss the quantum correlations.

\section{Quantum correlations required} For the anonymity tasks, we shall show that information is being hidden by the quantum correlations of the states. The main candidates are entanglement and discord, which can be defined mathematically by specifying forms for the correlated states. A bipartite state is entangled if it cannot be written in the separable form $\rho_{AB}=\sum_i p_i \rho_A^{(i)}\otimes\rho^{(i)}_B$.
A bipartite state is discordant if, for some local basis, it cannot be written in any of the three forms $\rho_{AB}=\sum_{i,j} p_{ij} |i\rangle\langle i|_A\otimes |j\rangle\langle j|_B,$ $\rho_{AB}=\sum_{i} p_{i} |i\rangle\langle i|_A\otimes \rho_{B|i},$ and $\rho_{AB}=\sum_{j} p_{j} \rho_{A|j}\otimes |j\rangle\langle j|_B,$  termed Classical-Classical (CC), Classical-Quantum (CQ), and Quantum-Classical (QC) respectively. Entangled states are a subset of discordant states.

\begin{figure}[h]
\centering
\includegraphics[scale=0.4]{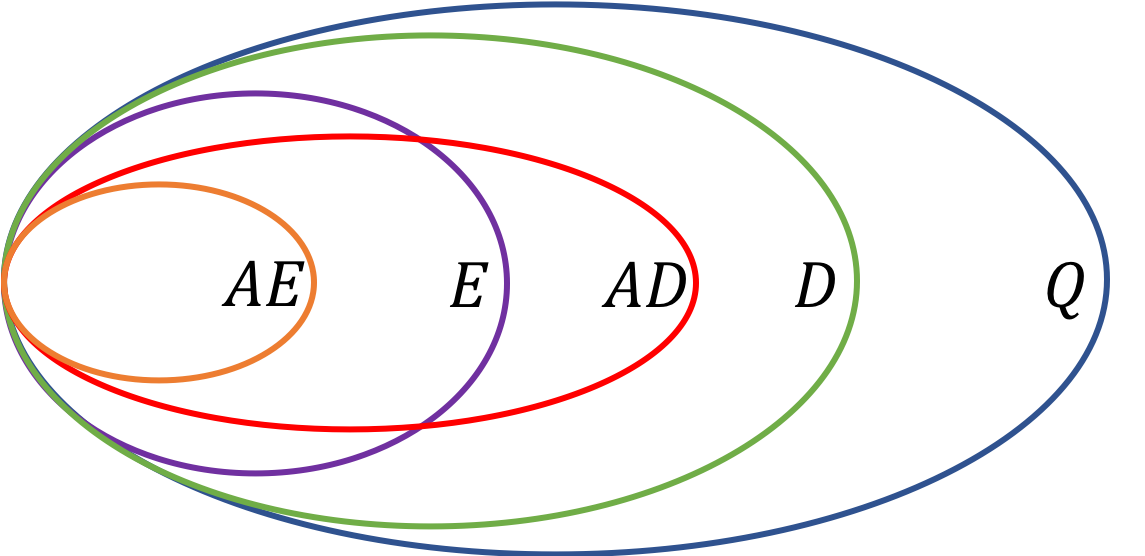}
\caption{Schematic summary of the relations between all quantum states (Q), and those that are discordant (D), aligned discordant (AD), entangled (E), and aligned entangled (AE).}
\label{SetRelationsFigure}
\end{figure}

As shown below, the WA and SA states form subsets of the known sets of correlated states. We therefore use the terms aligned discord and aligned entanglement for the WA and SA resources, respectively. For aligned discord we establish that discord is necessary but not sufficient, and entanglement is neither necessary nor sufficient. For aligned entanglement we show entanglement is necessary but not sufficient. This is illustrated in Fig.~\ref{SetRelationsFigure}.

\subsection{WA Hamiltonian condition} Before proving these results, we recast the WA conditions in terms of Hamiltonians, to describe families of unitaries that encode a continuous parameter. If we were to demand Eq.~(\ref{WACondition}) and (\ref{EncodingCondition}) without enforcing the continuous parameter requirement, then an anonymous encoding would be given by the classically correlated state $\rho_{AB}=\frac{1}{2}(|00\rangle\langle 00|+|11\rangle\langle 11|),$ with bit flip unitaries $U_A=\sigma^x_A,$ and $V_B=\sigma^x_B.$ Note that for the SA case there is no such distinction.

Writing $U_{A}(\theta)=e^{-i\theta H_{A}},$ and $V_{B}(\theta)=e^{-i\theta G_{B}},$ we see that the weak anonymity condition of Eq.~(\ref{WACondition}) is equivalent to requiring that there exist local Hermitian operators $H_{A},G_{B},$ for which
\begin{equation}
[H_A-G_B,\rho_{AB}]=0.
\label{AnonCommutatorCondition}
\end{equation}
Similarly the encoding condition of Eq.~(\ref{EncodingCondition}), becomes
\begin{equation}
[H_A,\rho_{AB}]\neq 0.
\label{EncodingCommutatorCondition}
\end{equation}
We can work with these conditions to intrinsically restrict to continuous parameter encodings.

\subsection{Aligned discord}
To prove discord is necessary we start with a CQ state $\rho_{AB}=\sum_{a}\lambda_{a}|\psi_{a}\rangle\langle\psi_{a}|\otimes\rho_{B|a},$ and take that for some choice of Hermitian operators $H_{A}$ and $G_{B}$ we have $[H_{A},\rho_{AB}]=[G_{B},\rho_{AB}].$ We project A onto $|\psi_{c}\rangle\langle\psi_{c}|$ and use the fact that $\langle\psi_{c}|[H_{A},|\psi_{a}\rangle\langle\psi_{a}|]|\psi_{c}\rangle=0$ to get $\lambda_{c}[G_{B},\rho_{B|c}]=0,$ $\forall c.$ From this we find $[H_{A},\rho_{AB}]=0.$ Hence we can only satisfy the anonymity condition if we violate the encoding condition. Essentially the same argument works for a QC state, so it is true for all non-discordant states. This proves that discord is necessary for WA states.

The fact entanglement is not necessary is proved with the Werner state~\cite{Werner1989} $\rho_W=a|\psi^{-}\rangle\langle\psi^{-}|+\frac{1-a}{4}\mathds{1}.$ The WA conditions can always be satisfied using $e^{ -i\theta |1\rangle\langle 1|},$ except when $a=1,$ but the state is not entangled for values of $a \leq \frac{1}{3}.$

To prove that neither entanglement nor discord are sufficient we use a two-qubit example. We construct a state that is entangled and discordant but is not of the appropriate form as given in Eq.~(\ref{WAStateForm}). To do this we first define $\rho_{1}=(\sqrt{a}|00\rangle+\sqrt{1-a}|11\rangle)(\sqrt{a}\langle00|+\sqrt{1-a}\langle11|)$ and $\rho_{2}=(\sqrt{b}|01\rangle+\sqrt{1-b}|10\rangle)(\sqrt{b}\langle01|+\sqrt{1-b}\langle10|).$ The example is formed by taking $\rho=m\rho_{1}+(1-m)\rho_{2}$. Picking the values $a=0.45,b=0.4,m=0.35,$ one can show that the resulting state is entangled (therefore discordant), but cannot satisfy the WA conditions (see Appendix \ref{appenDisAndEnWA} for details). Proof for discord alone can be given by the simpler example $\frac{1}{2}(|00\rangle\langle00|+|++\rangle\langle++|).$

\subsection{Aligned entanglement}
The fact entanglement is necessary for SA states follows from applying the Peres-Horodecki criterion~\cite{Peres1996,Horodecki1997} to states of the form presented in Eq.~(\ref{SAForm}). The fact that it is not sufficient also follows, since not all entangled states can be written in this form, for example the Werner state (see Appendix \ref{appenEntanglementNecForSA} for a detailed account). Note the Werner state also reveals that SA states are not simply the entangled WA states, but a strict subset of them. It also proves that steerability and Bell non-locality are not sufficient for aligned entanglement since for $a>\frac{1}{2}$ the state is steerable~\cite{Wiseman2007}, and for $a>\frac{1}{\sqrt{2}}$ it is Bell non-local~\cite{Horodecki1995}.

\section{Using non-ideal states}
\subsection{Robustness}\label{sec:NonIdealStates}
The anonymous metrology protocol has robust anonymity. If Alice and Bob verify that their state $\rho_{AB}$ is close to a WA/SA state $\sigma_{AB},$ in terms of trace distance $T(\rho_{AB},\sigma_{AB})\leq\epsilon,$ then this bounds Charlie's ability to correctly guess who applied the unitary. For the WA case we have $T(U_{A}\rho_{AB}U_{A}^{\dagger},V_{B}\rho_{AB}V_{B}^{\dagger})\leq 2\epsilon,$ and for the SA case we have $T(U_A|\psi\rangle_{ABC} ,V_B|\psi\rangle_{ABC}) \leq 2\sqrt{\epsilon-\epsilon^2},$ (see Appendix \ref{appenRobust} for derivations). This means that given a single copy, the probabilities for Charlie to correctly guess are bounded~\cite{Nielsen2010ch9}, as $P_{WA} \leq \frac{1}{2} + \epsilon,$ and $P_{SA} \leq \frac{1}{2} + \sqrt{\epsilon-\epsilon^2}.$ In general Alice and Bob send multiple copies, which Charlie could use to improve his guess. However using the property of the fidelity that $F(\rho_1^{\otimes n},\rho_2^{\otimes n})=F(\rho_1,\rho_2)^n,$ and the Fuchs-van de Graff inequality $1-F \leq T \leq \sqrt{1-F^2}$~\cite{Fuchs1999}, we find that $T(\rho_{1}^{\otimes n},\rho_{2}^{\otimes n}) \leq \sqrt{1-[1-T(\rho_1,\rho_2)]^{2n}}.$ Hence robustness for many copies follows.

\subsection{General figure of merit} Following similar considerations, we now define a general figure of merit for any bipartite state $\rho$ used for anonymous metrology. For given Hamiltonians $H,G$, we can bound the increase in Charlie's guessing probability to $\delta$ by limiting the number of copies sent to be no more than
\begin{equation}
	n_{\delta}=\frac{\log(1-(2\delta)^2)}{2\log \min_\theta F(\rho_H(\theta),\rho_G(\theta))},
\end{equation}
where $\rho_H(\theta)$ and $\rho_G(\theta)$ are the states Charlie is trying to discriminate between. The minimisation over $\theta$ ensures anonymity for the whole range of parameter values.

The usefulness of a state in parameter estimation may be quantified by the quantum Fisher information (QFI) $\mathcal{F}$~\cite{Petz2010}, which sets a lower limit on the uncertainty $\Delta \theta$ with which a parameter $\theta$ can be estimated, via the quantum Cram\'er-Rao bound: $\Delta \theta \geq (n \mathcal{F})^{-1/2}$ for $n$ measurements. For a unitary encoding $\rho_H(\theta) = U_A(\theta)\rho U_A^\dagger(\theta),$ the QFI is parameter-independent and we denote it as $\mathcal{F}(\rho;H_A).$ Since the QFI can depend on which party does the encoding, we define the average $\bar{\mathcal{F}}(\rho;H,G)=\frac{1}{2}(\mathcal{F}(\rho;H_A)+\mathcal{F}(\rho;G_B)).$

We combine $\bar{\mathcal{F}},$ and $n_{\delta}$ to form the figure of merit $n_{\delta}\bar{\mathcal{F}}.$ This captures the amount of parameter information that can be transferred to Charlie with $\delta$ anonymity. We identify the state-dependent part as the figure of merit
\begin{equation}
	M(\rho; H,G) = \frac{\bar{\mathcal{F}}(\rho;H,G)}{- \log \min_\theta F(\rho_H(\theta),\rho_G(\theta))}.
\end{equation}	
The larger $M(\rho; H,G)$, the better a state $\rho$ is for anonymous metrology with Hamiltonians $H,G$.

For a function purely of the state, we must maximize over all possible choices of Hamiltonian. In order for this to be well-defined, we need to bound the spectra -- therefore we define
\begin{equation}
	M(\rho) = \min_{H,G,\, \|H\| = \| G\| = 1 } M(\rho; H,G),
\end{equation}
where $\|\cdot \|$ is the operator norm.

\section{Conclusions} We have shown that quantum mechanics enables a metrology protocol whereby a continuous parameter may be determined whilst hiding the location where it was encoded. We established the nature of the quantum correlations responsible for this phenomenon, according to the level of privacy required. With a trusted source of states, discord is needed, while entanglement provides privacy with an untrusted source. The useful correlations have a particular symmetry, and are named aligned discord and entanglement respectively.

We note that the difference between the WA and SA tasks resembles device-dependent versus device-independent cryptography, where the former requires discord and the latter entanglement~\cite{Pirandola2014}. However the tasks are clearly distinct, and importantly we note that they do not produce the same sets of resourceful states. This is most readily seen by considering the aligned discord states that are useful for the WA task. For these states discord and entanglement are not sufficient, whereas for the device-dependent quantum key distribution task entanglement is sufficient. 

The fact the WA case only requires discord reduces the technological challenge in realizing protocols, with discord being relatively robust to noise~\cite{Werlang2009, Weedbrook2016}. The SA states are more practically challenging, but bring greater operational power. It is also noteworthy that there is a connection between aligned discord/entanglement and quantum coherence. This is indicated by the redefinition of the WA states in terms of modes of asymmetry given in Eq.~(\ref{WATwirling}). This suggests a potential link between the anonymity resources and the resource of quantum coherence~\cite{Baumgratz2014,Chitambar2016,Winter2016a,Yadin2016}. The anonymity resources should arguably be viewed as a hybrid of coherence and correlation. One could describe it as correlated coherence, though this appears distinct from the correlated coherence of~\cite{Kraft2018,Tan2016,Ma2018,Sun2017}.

Our results highlight an operational boundary within the hierarchy of quantum correlations, providing a novel nonclassical task whereby different correlations are at play depending on the desired level of anonymity.

\begin{acknowledgments}
The authors acknowledge discussions with Thomas Hebdige, and Hyukjoon Kwon. AP is funded by the EPSRC Centre for Doctoral Training in Controlled Quantum Dynamics, BY is funded by an EPSRC Doctoral Prize, and MSK by the Royal Society and Samsung GRO project.
\end{acknowledgments}

\bibliographystyle{unsrtnat}
\bibliography{anonmetreferences}

\onecolumn\newpage 
\appendix 

\section{Delocalised measurement}\label{appenNoDec}
Alice and Bob want to measure a system that has been put into a spatial superposition, without decohering it. We consider the unitary that they jointly perform. Alice sets up her lab such that if the particle comes to her then she performs the controlled unitary $U_A$, and Bob does similarly with $V_B$ (we leave the $\theta$-dependence implicit). Together this gives the full unitary as
\begin{equation}
W = |L\rangle_P \langle L|\otimes U_A \otimes \mathds{1}_B + |R\rangle_P\langle R|\otimes \mathds{1}_A\otimes V_B,
\end{equation}
We act with this on the initial state
\begin{equation}
\rho = (a|L\rangle+b|R\rangle)_P(a^*\langle L|+b^*\langle R|) \otimes\rho_{AB}.
\end{equation}
Writing the new state as a matrix in the $L,R$ basis we have
\begin{equation}
\rho' = 
\begin{bmatrix}
    |a|^2 U_A \rho_{AB} U^\dagger_A	   & a b^* U_A \rho_{AB}V^\dagger_B \\
    a^* b V_B \rho_{AB}U^\dagger_A       & |b|^2 V_B \rho_{AB}V^\dagger_B
\end{bmatrix}.
\end{equation}
We see that if 
\begin{equation}
\begin{aligned}
U_A \rho_{AB}U^\dagger_A = V_B \rho_{AB} V^\dagger_B = V_B \rho_{AB} U^\dagger_A ,
\end{aligned}
\label{TripleEquality}
\end{equation}
then we can factor out and write our state in the product form
\begin{equation}
\rho' = (a|L\rangle+b|R\rangle)_P(a^*\langle L|+b^*\langle R|)\otimes U_A \rho_{AB}U^\dagger_A.
\end{equation}
Factorisation is also possible with any phase factor on the right-hand side of Eq.~(\ref{TripleEquality}), resulting in a relative phase appearing in system $P$. This phase may be absorbed into the definition of $U_A$.

\section{SA condition with twirling superoperators}\label{appenSATwirl}
Starting from the SA condition $U_A(\theta)\rho_{AB}=V_B(\theta)\rho_{AB},$ we act from the right with $V_B^\dagger(\theta).$ We then multiply by $\frac{e^{-i\omega\theta}}{2\theta_0},$ integrate $\int_{-\theta_0}^{\theta_0} d\theta,$ and take the limit $\theta_0 \rightarrow \infty.$ Defining the split twirling operator
\begin{equation}
\mathcal{P}_{AB}^{\omega}(.)=\lim_{\theta_0 \rightarrow \infty}\frac{1}{2\theta_0}\int_{-\theta_0}^{\theta_0}d\theta e^{-i\omega \theta}U_A(\theta)(.)V^\dagger_B(\theta),
\end{equation}
we arrive at 
\begin{equation}\label{SATwirling}
\mathcal{P}_{AB}^{\omega}\rho_{AB}=\mathcal{P}_{B}^{\omega}\rho_{AB}.
\end{equation}
To go the other way we first note that by summing over $\omega$ we have $\sum_{\omega}\mathcal{P}_{AB}^{\omega}\rho_{AB}=\rho_{AB},$ even though $P_{AB}^\omega$ does not satisfy a completeness relation. This allows us to perform essentially the same argument as in the WA case. First we define the superoperator $\mathcal{W}_{AB,\theta}(.)=U_{A}(\theta)(.)V^{\dagger}_{B}(\theta),$ and note that $\mathcal{W}_{AB,\theta}\mathcal{P}_{AB}^{\omega}=e^{i\omega\theta}\mathcal{P}_{AB}^{\omega}.$ We then act on Eq. (\ref{SATwirling}) with $\mathcal{W}_{AB,\theta}\mathcal{V}_{B,\theta},$ cancel the $e^{i\omega\theta}$ terms and sum over $\omega$ to arrive back at the original SA conditions. However, the split twirling operator is not an established tool, and it does not appear straightforward to go from these expressions to a form for the useful states.

\section{Form for multipartite states}\label{appenMultiPart}
\subsection{WA case}
To generalise the WA states is straightforward. We consider the case of $n$ subsystems labeled $1,2,...,n$ and demand that the anonymity condition holds between every pair of subsystems. Now the WA condition in terms of the twirling superoperators becomes
\begin{equation}
\mathcal{P}_{\alpha}^{\omega}\rho_{12...n} = \mathcal{P}_{\beta}^{\omega}\rho_{12...n},
\end{equation}
for all $\alpha,\beta$ in the set ${1,2,...,n}.$ In direct analogy to the bipartite case we have
\begin{equation}
\rho^{(\omega_{1},...,\omega_{n})}_{1...n}=\sum_{\substack{i,i',...,k,k',\\ E_{i'}=E_{i}+ \omega_{1},..., \\ E_{k'}=E_{k}+ \omega_{n}}}c_{ii'...kk'} |i\rangle_1\langle i'|\otimes...\otimes |k\rangle_n\langle k'|,
\end{equation}
and now we can write the WA states as
\begin{equation}
\rho_{12...n}=\sum_\omega \rho^{(\omega,\omega,...,\omega)}_{12...n},
\end{equation}
where as before we require non-zero terms for $\omega\neq 0$ so that encoding is possible.

\subsection{SA case}
For the SA state we generalise the condition to be $U_{\alpha}\rho_{12...n}=U_{\beta}\rho_{12...n},$ for all $\alpha,\beta$ in the set ${1,2,...,n}.$ We apply the same approach as before, taking matrix elements in the eigenbasis of the unitaries. Working in the non-degenerate case with the same approach as before we find each of these equations gives us an expression for the allowed non-zero terms like $|iikl...n\rangle\langle jjk'l',,,n|,$ and $|ikil...n\rangle\langle jk'jl',,,n|.$ Taking all such conditions together we see the only non-zero terms allowed are of the form $|iii...i\rangle\langle jjj...j|.$ And so we have the form
\begin{equation}
    \rho_{12...n}=\sum_{i,j}\rho_{ij}|ii...i\rangle\langle jj...j|.
\end{equation}
One can then extend to allow for degeneracy in the same way as before by introducing degeneracy labels.

\section{Proof discord and entanglement not sufficient for aligned discord}\label{appenDisAndEnWA}

Here we prove by examples that discord and entanglement are not sufficient for a state to be a WA state. The proof uses the explicit forms of WA states for two qubits, so we first discuss this.

For a two qubit bipartite state $\rho_{AB},$ where the eigenvalues of $\rho_A=\text{Tr}_B(\rho_{AB}),$ are non-degenerate and similarly for $\rho_B,$ such that they have unique eigenvectors, then the WA states are those that, when written in the eigenbasis of their reduced density matrices, have the form

\begin{equation}\label{QubitsMatrixForm}
\rho =
\begin{bmatrix}
    \alpha      & 0 & 0 & \epsilon \\
    0      & \beta & 0 & 0\\
    0      & 0 & \gamma & 0 \\
    \epsilon^*      & 0 & 0 & \delta
\end{bmatrix}\text{or}
\begin{bmatrix}
    \alpha      & 0 & 0 & 0 \\
    0      & \beta & \epsilon & 0\\
    0      & \epsilon^* & \gamma & 0 \\
    0     & 0 & 0 & \delta
\end{bmatrix},
\end{equation}

where $\epsilon\neq0$. (These two forms are related by a relabelling of the eigenbasis of $\rho_A$.)

We can arrive at this by considering the general form given in Eq. (\ref{WATwirling}), however here we present a constructive proof from the Hamiltonian WA conditions of Eq. (\ref{AnonCommutatorCondition}) and (\ref{EncodingCommutatorCondition}). First note the local Hamiltonian $H_A$ must share an eigenbasis with $\rho_A$ and similarly for $G_B.$ Using this local eigenbasis we take matrix elements of Eq. (\ref{AnonCommutatorCondition}) to get $(h_{i}-g_{j}-h_{i'}+g_{j'})\langle ij|\rho_{AB}|i'j'\rangle=0,$ where $h,g$ are  the local Hamiltonian eigenvalues.
From this equation we see that the diagonal terms $\langle ij|\rho_{AB}|ij\rangle$ are unconstrained. To see what other terms are free to be non-zero we need to consider when we can make $(h_{i}-g_{j}-h_{i'}+g_{j'})=0.$

Since we have 2 qubits we have 4 eigenvalues to set: $h_0,h_1,g_0,g_1.$ The encoding condition Eq. (\ref{EncodingCommutatorCondition}) enforces $h_0\neq h_1$ and $g_0\neq g_1,$ and that $\rho_{AB}$ has at least one off-diagonal term, since $H_A\otimes \mathds{1}_B$ is diagonal and diagonal matrices commute with each other.  We now have two options, choose $h_0=g_0,$ and $h_1=g_1,$ or $h_0=g_1,$ and $h_1=g_0.$ The first case allows the terms $\langle 00|\rho_{AB}|11\rangle,$ and $\langle 11|\rho_{AB}|00\rangle,$ to be non-zero and the second case allows $\langle 01|\rho_{AB}|10\rangle,$ and $\langle 10|\rho_{AB}|01\rangle.$
Putting this all together we arrive at the forms stated in Eq. (\ref{QubitsMatrixForm}).

The facts that entanglement and discord are not sufficient are proved by a two qubit example that is not of the form given in Eq. (\ref{QubitsMatrixForm}) but is entangled (and therefore discordant).
First we define $\rho_{1}=(\sqrt{a}|00\rangle+\sqrt{1-a}|11\rangle)(\sqrt{a}\langle00|+\sqrt{1-a}\langle11|)$ and $\rho_{2}=(\sqrt{b}|01\rangle+\sqrt{1-b}|10\rangle)(\sqrt{b}\langle01|+\sqrt{1-b}\langle10|).$ Now we can form $\rho=m\rho_{1}+(1-m)\rho_{2}$. Choosing $a,b\neq\frac{1}{2}$ ensures we have unique local eigenvectors. Picking the fairly arbitrary values $a=0.45,b=0.4,m=0.35,$ we find the matrix in its local eigenbasis is approximately

\begin{equation}
\begin{bmatrix}
    0.2     & 0 & 0 & 0.2 \\
    0      & 0.3 & 0.3 & 0\\
    0      & 0.3 & 0.4 & 0 \\
    0.2     & 0 & 0 & 0.1
\end{bmatrix},
\end{equation}
where we are quoting values only to one significant figure. This is not in one of the viable forms given in Eq. (\ref{QubitsMatrixForm}). Taking the partial transpose it has a negative eigenvalue. Thus by the Peres-Horodecki criterion \cite{Peres1996,Horodecki1997} the state is entangled.
Since entangled states are always discordant, this example proves neither discord nor entanglement are sufficient. However if one wanted to show it just for discord, the simpler example $\frac{1}{2}(|00\rangle\langle00|+|++\rangle\langle++|),$ suffices. This concludes the proof.

For completeness we present an alternative way to arrive at Eq. (\ref{QubitsMatrixForm}), using tools from asymmetry theory \cite{Bartlett2007}. We can write the WA anonymity condition as a symmetry constraint by using the G-twirling superoperator. We have

\begin{equation}
\mathcal{G}(\rho_{AB})=\rho_{AB},
\end{equation}
where we define
\begin{equation}
\mathcal{G}(\rho_{AB}) = \lim_{\theta_0 \to \infty} \frac{1}{2 \theta_0} \int_{-\theta_0}^{\theta_0} d\theta \; \mathcal{U}_{A,\theta}  \otimes \mathcal{V}_{B,\theta}^\dagger \rho_{AB}.
\end{equation}
Taking the two-qubit case we write
\begin{equation}
U_{A}(\theta) \otimes V_{B}(\theta)^\dagger =
\begin{bmatrix}
    1      & 0 \\
    0      & e^{i a \theta} \\
    
\end{bmatrix}\otimes
\begin{bmatrix}
    1      & 0 \\
    0      & e^{-i b \theta} \\
    
\end{bmatrix} = 
\begin{bmatrix}
    1     & 0 & 0 & 0 \\
    0      & e^{-i b \theta} & 0 & 0\\
    0      & 0 & e^{i a \theta} & 0 \\
    0     & 0 & 0 & e^{i (a-b) \theta}
\end{bmatrix}.
\end{equation}
We then find
\begin{equation}
\mathcal{U}_{A,\theta} \otimes \mathcal{V}^{\dagger}_{B,\theta}\rho_{AB} = 
\begin{bmatrix}
    1     & e^{i b \theta} & e^{-i a \theta} & e^{-i (a-b) \theta} \\
    
    e^{-i b \theta}      & 1 & e^{-i (a+b) \theta} & e^{-i a \theta}\\
    
    e^{i a \theta}      & e^{i (a+b) \theta} & 1 & e^{i b \theta} \\
    
    e^{i (a-b) \theta}     & e^{i a \theta} & e^{-i b \theta} & 1
\end{bmatrix} * \rho_{AB},
\end{equation}
where $*$ denotes the entrywise product in the computational basis. When we integrate to perform the G-twirling, the two choices of either $a=b,$ or $a=-b,$ give the two forms of viable density matrix, as in Eq. (\ref{QubitsMatrixForm}).

\section{Proof entanglement necessary but insufficient for aligned entanglement}\label{appenEntanglementNecForSA}

Starting from the form given in Eq. (\ref{SAForm}) we find that entanglement is necessary for aligned entanglement. To see this we write $\rho_{AB}=\sum_{i,j}\alpha_{ij}|i i\rangle_{AB}\langle j j|,$ where we have dropped the degeneracy labels. In order to not violate the encoding condition of Eq. (\ref{EncodingCondition}) we must have a $k$ and $l$ for which $k\neq l,$ and $\alpha_{kl}=e^{i\phi_{kl}}|\alpha_{kl}|\neq 0.$ We can apply the local unitary $e^{-i\phi_{kl}|k\rangle_{A}\langle k|},$ without affecting the entanglement, and we absorb the phases into the $\alpha_{kl},$ and $\alpha_{lk},$  such that now they are both real and positive. We then perform the partial transpose to get $\rho_{AB}^{T_B}=\sum_{i,j}\alpha_{ij}|i j\rangle_{AB}\langle j i|,$ and see that the state $|kl\rangle-|lk\rangle,$ is an eigenvector, with the negative eigenvalue $-|\alpha_{kl}|.$ Hence by the Peres-Horodecki criterion \cite{Peres1996,Horodecki1997}, aligned entangled states are always entangled. Below we also give an alternative proof that works from the SA conditions without making use of the form from Eq. (\ref{SAForm}).

To prove that entanglement is not sufficient we use the Werner state example $\rho=a|\psi^{-}\rangle\langle\psi^{-}|+\frac{1-a}{4}\mathds{1},$ and show the only way this can satisfy the strong anonymity condition of Eq. (\ref{SACondition}) is if it violates the encoding condition. The anonymity condition becomes $aU_{A}(\theta)|\psi^{-}\rangle\langle\psi^{-}|+\frac{1-a}{4}U_{A}(\theta)=a V_{B}(\theta)|\psi^{-}\rangle\langle\psi^{-}|+\frac{1-a}{4} V_{B}(\theta).$
We now act from the right with $|\psi^{-}\rangle,$ and we get through to
$U_{A}(\theta)|\psi^{-}\rangle=V_{B}(\theta)|\psi^{-}\rangle.$
Substituting this into the original condition we have
$U_{A}(\theta)\otimes \mathds{1}_{B}=I_{A}\otimes V_{B}(\theta).$
This implies $U_{A}(\theta)=\mathds{1}_{A},$ and hence we must violate the encoding condition. This is true for $0<a<1,$ and thus is true for values of $a$ for which the state is entangled. Hence entanglement is not sufficient for aligned entanglement.

Now we present a proof that entanglement is necessary for SA states without making explicit use of the form given in Eq. (\ref{SAForm}). First we state and prove a useful lemma.

Given a probability distribution $\{p_{j}\},$ and a set of complex numbers $\{Z_{j}\},$ where $|Z_{j}|\leq1.$ Then $\sum_{j}p_{j}Z_{j}=e^{i\xi},$ iff $Z_{j}=e^{i\xi},$ $\forall j$ where we only consider $j$ for which $p_{j}\neq0.$

To see this is true, note any complex number on the boundary of the unit disk cannot be expressed as a convex combination of other complex numbers in the unit disk. This should be convincing but for completeness we give a more formal proof.

For a set of non-zero complex numbers $\zeta_j$, we have $|\sum_j\zeta_j|\leq\sum_j |\zeta_j|,$ with equality iff $\text{arg}(\zeta_j)=\text{arg}(\zeta_k), \forall j,k.$ This is just a restatement of the polygon inequality (generalisation of the triangle inequality) for complex numbers. We now write $\sum_{j}p_{j}Z_{j}=\sum_j\zeta_j.$ Now we have $\sum_j |\zeta_j|=\sum_j p_j|Z_j|\leq\sum_j p_j= 1,$ where we used $|Z_j|\leq 1,$ so the equality holds when $|Z_j|=1 \forall j.$ We now have $|\sum_j\zeta_j|\leq\sum_j |\zeta_j|\leq 1,$ but $|\sum_j\zeta_j|=|e^{i\xi}|=1,$ and so the three terms are all equal. The conditions on the inequalitites then tell us every $Z_j$ has unit magnitude and the same argument and we see this argument must be $\xi$ giving us $Z_{j}=e^{i\xi}.$ The converse is trivial.

We can now prove entanglement is necessary for SA states. We first obtain from anonymity condition $\text{Tr}(U_{A}(\theta)V_{B}^{\dagger}(\theta)\rho_{AB})=1.$ With a separable state $\rho_{AB}=\sum_{j}p_{j}\rho_{A}^{(j)}\otimes\rho_{B}^{(j)},$ the condition becomes $\sum_{j}p_{j}\text{Tr}(U_{A}(\theta)\rho_{A}^{(j)})\text{Tr}(V_{B}^{\dagger}(\theta)\rho_{B}^{(j)})=1.$ We have $|\text{Tr}(U_{A}(\theta)\rho_{A}^{(j)})|\leq1$, and $|\text{Tr}(V_{B}(\theta)\rho_{B}^{(j)})|\leq1$, via the above lemma we have $\text{Tr}(U_{A}(\theta)\rho_{A}^{(j)})\text{Tr}(V_{B}^{\dagger}(\theta)\rho_{B}^{(j)})=1$. This can only be true if $|\text{Tr}(U_{A}(\theta)\rho_{A}^{(j)})|=1,$ so $\text{Tr}(U_{A}(\theta)\rho_{A}^{(j)})=e^{i\phi(j)}.$ Using the eigendecomposition $\rho_{A}^{(j)}=\sum_{k}q_{k}^{(j)}|\psi_{k}^{(j)}\rangle\langle\psi_{k}^{(j)}|,$ this becomes $\sum_{k}q_{k}^{(j)}\langle\psi_{k}^{(j)}|U_{A}(\theta)|\psi_{k}^{(j)}\rangle=e^{i\phi_{A}(j)}$. Again using the lemma we have $\langle\psi_{k}^{(j)}|U_{A}(\theta)|\psi_{k}^{(j)}\rangle=e^{i\phi_{A}(j)}$. This implies $U_{A}(\theta)|\psi_{k}^{(j)}\rangle=e^{i\phi_{A}(j)}|\psi_{k}^{(j)}\rangle.$ We now see that this violates the encoding condition since $U_A(\theta)\rho_{AB}U^\dagger_A(\theta)=\sum_{j,k}p_{j}q_{k}^{(j)}e^{i\phi_{A}(j)}|\psi_{k}^{(j)}\rangle\langle\psi_{k}^{(j)}|e^{-i\phi_{A}(j)}\otimes\rho_{B}^{(j)}=\rho_{AB}.$ We have therefore shown that all separable states cannot satisfy the SA conditions and hence entanglement is necessary.

\section{Error and robustness}\label{appenRobust}
\subsection{WA case}
Alice and Bob use a selection of the copies they receive to verify that the state they are using $\rho_{AB}$ is close, in terms of trace distance, to a WA state $\sigma_{AB}.$ Formally we say they verify that
\begin{equation}\label{ABTraceDist}
T(\rho_{AB},\sigma_{AB})\leq\epsilon.
\end{equation}
We now show that this leads to a bound on $T(U_{A}\rho_{AB}U_{A}^{\dagger},V_{B}\rho_{AB}V_{B}^{\dagger})$. This quantifies Charlie's ability to distinguish whether Alice or Bob applied their unitary, since the maximal probability of correctly guessing the state is $P = (1 + T(U_A \rho_{AB} U_A^\dagger, V_A \rho_{AB} V_A^\dagger))/2$  \cite{Nielsen2010ch9}.

Starting from Eq. (\ref{ABTraceDist}) we use the fact that trace distance is preserved under unitaries to write
\begin{equation}
T(U_{A}\rho_{AB}U_{A}^{\dagger},\sigma'_{AB})\leq\epsilon,
\end{equation}
\begin{equation}
T(V_{B}\rho_{AB}V_{B}^{\dagger},\sigma'_{AB})\leq\epsilon,
\end{equation}
where we have defined $\sigma'_{AB}=U_{A}\sigma_{AB}U_{A}^{\dagger}=V_{B}\sigma_{AB}V_{B}^{\dagger},$ using the fact $\sigma_{AB}$ is a WA state. We now use the triangle inequality $T(A,C)\leq T(A,B)+T(B,C),$ to arrive at
\begin{equation}
T(U_{A}\rho_{AB}U_{A}^{\dagger},V_{B}\rho_{AB}V_{B}^{\dagger})\leq 2\epsilon.
\end{equation}

\subsection{SA case}
Again Alice and Bob use some of their states to establish Eq. (\ref{ABTraceDist}). However, for the SA case we need to consider distinguishability for the fully purified states, so we need to bound $T(U_{A}(\theta)|\psi\rangle_{ABC} ,V_{B}(\theta)|\psi\rangle_{ABC}).$

First consider the fidelity between the two states $U_{A}|\psi\rangle_{ABC},$ and $V_{B}|\psi\rangle_{ABC}.$ This fidelity is given by 
\begin{equation}\label{FidelityABC}
F(U_{A}|\psi\rangle_{ABC},V_{B}|\psi\rangle_{ABC}) = |\langle\psi|_{ABC}V_{B}^{\dagger} U_{A}|\psi\rangle_{ABC}|=|\text{Tr}(V_{B}^{\dagger}U_{A}\rho_{AB}))|.
\end{equation}
Now consider
\begin{eqnarray*}
1-|\text{Tr}(V_{B}^{\dagger}U_{A}\rho_{AB})| & \leq & |1 - \text{Tr}(V_{B}^{\dagger}U_{A}\rho_{AB})|,  \\
& = & |\text{Tr}(\sigma_{AB}-V_{B}^{\dagger}U_{A}\rho_{AB})|,  \\
& = & |\text{Tr}(V_{B}^{\dagger}U_{A}(\sigma_{AB}-\rho_{AB}))|,  \\
& \leq & \text{Tr}(|V_{B}^{\dagger}U_{A}(\sigma_{AB}-\rho_{AB})|),   \\
& = & \text{Tr}(|\sigma_{AB}-\rho_{AB}|),  \\
& = & 2 T(\rho_{AB},\sigma_{AB}),
\end{eqnarray*}
where in the third line we used that fact that $\sigma_{AB}$ is an SA state. From this we have $|\text{Tr}(V_{B}^{\dagger}U_{A}\rho_{AB})|\geq 1-2 T(\rho_{AB},\sigma_{AB})\geq 1-2\epsilon,$ and using this with Eq. (\ref{FidelityABC}) leads to a bound on the fidelity of
\begin{equation}
F(U_{A}|\psi\rangle_{ABC},V_{B}|\psi\rangle_{ABC}) \geq 1-2\epsilon.
\end{equation}
We now change this to an inequality in terms of the trace distance by using the fact that for pure states $T(\psi,\phi)=\sqrt{1-F(\psi,\phi)^2},$ to arrive at 
\begin{equation}
T(U_A|\psi\rangle_{ABC} ,V_B|\psi\rangle_{ABC}) \leq 2\sqrt{\epsilon-\epsilon^2}.
\end{equation}

\end{document}